- Title: Innovation for Sustainability in the Global South: Bibliometric findings from management & business and STEM (Science, Technology, Engineering and Mathematics) fields in developing countries

- Authors:

Julián David Cortés-Sánchez. Corresponding author.

Principal Professor – School of Management and Business

Universidad Del Rosario, Colombia

Invited Researcher – Fudan Development Institute

Fudan University, China

Email: julian.cortess@urosario.edu.co

Mireia Guix

Lecturer – School of Business

University of Queensland, Australia

Email: m.guix@uq.edu.au

Assistant Professor – School of Management and Business

Universidad Del Rosario, Colombia

Email: mireia.guix@urosario.edu.co

Katerina Bohle Carbonell

Post-Doctoral Fellow – Whitaker Institute

National University of Ireland Galway, Ireland

Email: katerina.bohle-carbonell@nuigalway.ie

Ronin Institute, United States

Email: katerina.bc@ronininstitute.org

# Innovation for Sustainability in the Global South

## Bibliometric findings from management & business and STEM (Science, Technology, Engineering and Mathematics) fields in developing countries

**Abstract**

Research on innovation and sustainability is prolific but fragmented. This study integrates the research on innovation in management and business and STEM fields for sustainability in a unified framework for the case of developing countries (i.e., the Global South). It presents and discusses the output, impact, and structure of such research based on a sample of 14,000+ articles and conference proceedings extracted from the bibliographic database Scopus. The findings reveal research output inflections after global announcements such as UN-Earth Summits. The study also reveals the indisputable leadership of China in overall output and research agenda-setting. Nonetheless, countries such as India, Mexico, and Nigeria are either more efficient or impactful. GS research published in highly reputable journals is still scarce but increasing modestly. Central topic clusters (e.g., *knowledge management*) remain peripheral to the global Sustainable Development Goals (SDGs) research landscape. Finally, academic-corporate collaboration is in its infancy and limited to particular economic sectors: energy, pharmaceuticals, and high-tech.

## 1 Introduction

Research on the topics of innovation and sustainability has been prolific. A search in the bibliographic database Scopus found 107,000 and 49,000 results, respectively. Despite the volume of research, details on the intersection between these fields are lacking, and more so in the context of developing countries (i.e., Global South: GS). Research production and appropriation from and to the GS in both fields are crucial for the global development agenda. Ten of the 17 Sustainable Development Goals (SDGs) mention either innovation or technology, highlighting their transversal role in the achievement of sustainable development (UNDP, 2019; United Nations, 2018, 2017).

Research within the GS, particularly Asia, has grown consistently. China is the world's largest scientific articles producer –661,000+ in 2020– and ranked second in R&D investment after the United States (Scopus, 2021; Tollefson, 2018). However, China might be an outlier in the GS. Innovation-related research performance in East Asia, Latin America and the Caribbean (LAC), and Africa shows them at a disadvantage compared with developed countries (Merigó et al., 2016). Also, the GS is falling short of the 1.68% GDP world average in R&D investment (United Nations, 2019). The upside is that there is are high expectations for increased research output from the GS in the forthcoming years (Finardi, 2015).

Bibliometrics can be used to explore the substantial volume of research on innovation and sustainability systematically. Bibliometrics is the quantitative understanding of volume, impact, and structure of scientific and technological literature, such as scholarly communications and patents (Fairthorne, 2005; Pritchard, 1969). Bibliometric appraisals reveal trending or converging research topics and their impact (Caviggioli, 2016; Chao et al., 2007); the social capital of authors, academic units, institutions, or regions (Tuzi, 2005); among others (Zupic and Čater, 2015).

Research dissecting the study of innovation and/or sustainability on the GS using bibliometric techniques has produced substantial insights on the SDGs, sustainability science, and their connection to management and a mix of disconnected topics, ranging from industrial relations to supply chain management. Despite these developments, several questions remain unexplored, such as what is the production dynamic? What is its impact? Are there concentrated efforts on specific research topics? How can we characterize and categorize the research in terms of topics, fields and co-authorship? Consequently, this study aims to conduct a comprehensive bibliometric outlook on the output, impact, and structure of the research on innovation for sustainability in the GS (henceforth: iS-GS).

The contributions of this study are threefold. Modern innovation studies have clustered around three disciplinary perspectives, namely the economics of R&D, organizing innovation, and innovation systems (Fagerberg et al., 2012). This study delves into an integrated framework of iS-GS, including research from business and management and STEM (Science Technology Engineering and Mathematics) fields.

Bearing in mind that local-global insights are vital inputs for strategic policy-making (Boulanger and Bréchet, 2005), this study covers three continents (i.e., Asia, Africa, and LAC) yet discusses national-level features. Finally, the dataset used is available in open access for academics and practitioners to conduct detailed analysis, replications, or triangulation, as requested by the academic community (Munafò and Davey Smith, 2018; Servick, 2018).

This paper proceeds as follows: in the next section, the literature review is presented. Next, the methodology section describes the data sources and the bibliometric methods used. That is followed by the results, which present the main findings on output and impact by type of document 1996-2018; documents/average researchers in R&D by continent and country; most productive institutions; citations/document ratio by continent and country; and publications by groups (exceptional, high, middle-low impact). Regarding structure, the following section presents the bibliographic coupling network; the co-authorship network by country; and the academic-corporate collaboration. The discussion and conclusion section places the findings within a broader body of literature and addresses study limitations and further research.

## 2 Literature review

A definition of innovation or sustainability research depends on the context (Brown et al., 1987). In this study, innovation research is defined as the systematic study of inventions, improvements, and implementations of new management practices, processes, structures, or techniques for achieving an organization's goals (Birkinshaw et al., 2008). However, sustainability research, as defined by the sustainable development agenda, is the study of "[the development that] meets the needs of the present generation without compromising the ability of future generations to meet their own needs" (World Commission on Environment and Development, 1987, p. 41).

Bibliometric studies on innovation from the GS are policymaking-oriented in some countries, notably China, and topic-diverse in other regions, such as LAC and Africa. The drive for bibliometric studies in China is their use in developing STi policy in knowledge-intensive and high-tech sectors (e.g., bio and nanotechnologies, pharmaceutics or renewable energy) (Guan and He, 2007; Huang et al., 2014; Li et al., 2017, 2015). In developing regions such as LAC and Africa, research topics range from *Schumpeterian* innovation and cooperation (Lazzarotti et al., 2011; Lopes and De Carvalho, 2012) to social innovation (Silveira and Zilber, 2017), passing through innovativeness measures (De Carvalho et al., 2017), industry relations (Manjarrez et al., 2016), business models (Ceretta et al., 2016), financing on innovation (Padilla-Ospina et al., 2018) and supply chain management (Tanco et al., 2018). The use of bibliometric insights for public policy in South Africa has focused on concrete sectors and fields (i.e., biotechnology and energy) (Makhoba and Pouris, 2016). Few studies have produced regional or subcontinental outlooks (Cortés-Sánchez, 2019; Pouris and Pouris, 2009).

Research on sustainability has grown nearly exponentially and increasingly geographically diverse since the 1980s, with an emphasis on the management of human, social, and ecological systems (Bettencourt and Kaur, 2011). Further research clustered around 15 topics (e.g., agriculture, fisheries, ecological economics) and assembled a framework on indicators setting, measurement methods, and causal chain analysis, among others (Kajikawa, 2008; Kajikawa et al., 2007). Recent findings from comprehensive bibliometric studies focused on the SDGs stated that there are two main domains (i.e., cluster): *health and healthcare*, and *environment, agriculture and sustainability science,* which are connected by *water supply and sanitation* cluster (Nakamura et al., 2019)*.*

In examining the relationship between research on SDGs and business, management and innovation, Jia et al. (2019) found a shift from focusing exclusively on economic growth and consumption toward an integrative framework that also addresses social development and environmental protection. Cui and Zhang's (2018) study on the circular economy also connected innovation and sustainability. They found that China's 13th Five-Year Plan aimed to work towards an ecological civilization and the comprehensive construction of a well-off society, which produced interactive feedback between public policy and academic research. Vatananan-Thesenvitz et al. (2019) produced several findings, some of which identified highly productive countries and highly co-authorship dynamics (e.g., USA, UK, Canada, Australia, New Zealand, Netherlands), and highly co-occurring keywords (e.g., planning, education, environmental protection/management and decision making).

In sum, two bibliometric pathways of literature stand out: i) the SDGs, sustainability science and their connection to business and management; and ii) a mix of disconnected topics, ranging from industrial relations to supply chain management. This study extends these efforts by integrating the research on innovation from business and management and STEM fields for sustainability in the GS.

## 3 Methodology

### 3.1 Data

Most of the studies reviewed used bibliographic data from Web of Science (WoS) (Kajikawa et al., 2007; Nakamura et al., 2019). In contrast, this study used Elsevier's Scopus because Scopus has broader journal coverage and higher social sciences coverage than WoS (Gavel & Iselid, 2008; Mongeon & Paul-Hus, 2016). Fig. 1 contains the detailed query.

Keywords for the search relied on key-term definitions from two reviews on innovation (Baregheh et al., 2009) and sustainability (Glavič and Lukman, 2007). In the first one, the authors conducted an SLR (Systematic Literature Review) of definitions on innovation (i.e., types and processes) in reputable organizational studies journals (e.g., *Management Science, Organization Science*) (1964-2007) followed by a content analysis similar to that of *ethnographic content analysis* (i.e., letting categories to emerge by counting word frequencies). In the second one, the authors surveyed 51 terms from the United Nations Environmental Program, the OECD, the *Journal of Cleaner Production*, and conducted a content analysis to improve their definition.

For the Scopus search query, we use the Booleans "OR" for keywords within innovation and sustainability, assembling two sets and "AND" unifying both sets. For instance, any document with the keywords *new* OR *innovation* "AND" *pollution control* OR *zero waste*, was sourced. The search was limited to document titles to ensure both topics' centrality in each document (Nakamura et al., 2019). Titles are the primary introduction to readers and reviewers by catching attention, describing the contents, and calling to read further. Also, titles often contain 2 of the most relevant indexing keywords. On the other hand, abstracts aim to explain *what the* *research is about; what methods have been used; what was found out,* among other questions (Springer, n.d.; Taylor and Francis, n.d.). Thus, to catch a non-specialized audience's attention, broader and peripheral topics are often mentioned in the abstract, increasing the odds of selections with no centrality of such topics for the research.

The query was restricted to institutions located in the GS: 103 countries in total, including all countries in LAC, Africa, and Asia except for Hong Kong, Japan, South Korea, Macau, Singapore, and Taiwan (International Telecommunications Union - ITU, 2018; Wikipedia, 2015). The query was restricted to articles and conference paper(s) (CP) since CP are vital in STEM scholarly communication (Lisée et al., 2008). The documents considered were those related to business and management subjects noted by Scopus' journal subjects (SCImago, 2018). The documents on STEM fields/disciplines considered were derived from a cross-checking assessment conducted by the authors. The authors individually and collectively assessed the STEM-approved fields by the National Science Foundation (2014) and the list of Scopus' journal subjects. If a field consistently appeared in both lists, the journals/CP in such a field were considered. The search was limited to 1996-2018, considering Scopus restrictions regarding documents indexation previous to 1996 (Ioannidis et al., 2016). Scopus found 14,900+ documents: 9,650+ articles and 5,320+ CP (Scopus, 2021).The following permanent link grants access to the dataset and the complete query (*space for posting permanent link after review or if reviewers ask for it*).

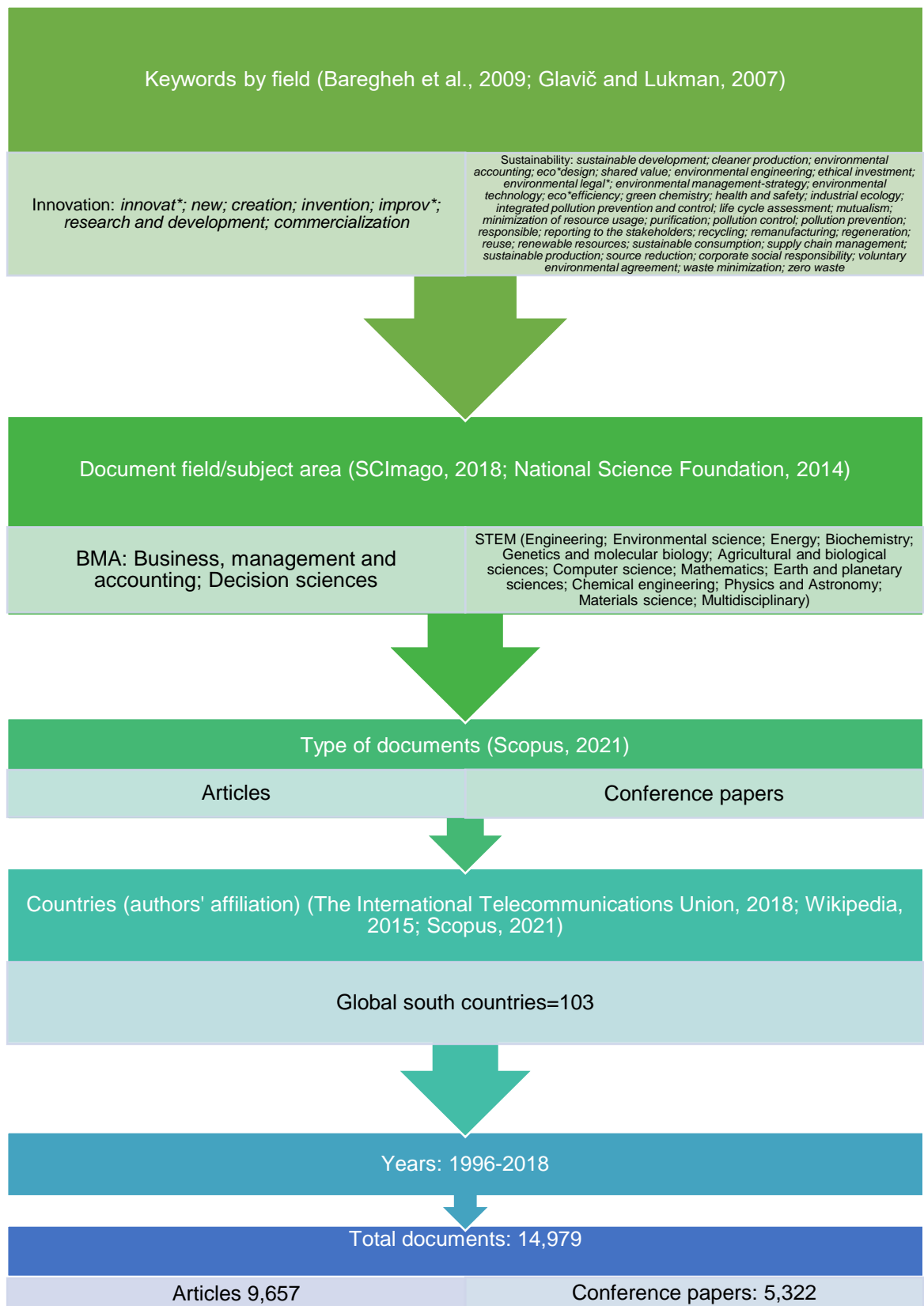

**Figure 1** Source: the authors, based on Cortés-Sánchez et al. (2020) Baregheh et al. (2009); Glavič & Lukman (2007); SCImago (2018); National Science Foundation (2014); The International Telecommunications Union (2018); Wikipedia (2015); and Scopus (2021).

## 3.2  Methods

The production and impact sections are descriptive. Regarding the structure section, Yan and Ding (2012) assessed multiple bibliometric networks to identify both similarities and differences. They found that coword and co-citation networks are cognitive and citation-based; citation and bibliographic coupling networks (BCN) are social and citation-based; co-authorship networks (CAN) are social-based networks; and topic networks are cognitive-based. Here we used BCN and CAN at the national level since few studies have produced regional or subcontinental outlooks (Cortés-Sánchez, 2019; Pouris and Pouris, 2009), and previous studies have presented related results using co-citation and co-word networks (Vatananan-Thesenvitz et al., 2019).

The network structure between publications highlights how articles are related to each other by shared references (Small, 1973). The BCN is used to both predict and describe emergent topics (Small, Boyack, & Klavans, 2014; Boyack & Klavans, 2010). A BCN computes the shared references between two documents (Zupic & Cater, 2015). For instance, in Fig. 9, if two circles (i.e., nodes) are connected (i.e., via a link), it means that both documents cited at least one document in common. Therefore, clusters are formed based on shared knowledge. BCN considers both mature and new literature (Zupic & Cater, 2015). Cluster analysis (edge betweenness), central publications (degree centrality), and bridging publications (betweenness centrality) techniques applied reveal emergent topics. On the other hand, CAN is one of the most tangible and reliable methods to study scientific collaboration and the creation of social networks of researchers, departments, institutions, and countries by linking co-authors' institutional/country affiliation in a given document (Glänzel and Schubert, 2005).

## 4  Result analysis

Scopus' SciVal data for the period 2009-2018 reported the involvement of 43,166 authors, who collectively published 12,296 documents and generated 90,403 citations. Fig. 2 shows the output by type of document. The average annual growth rate was 18% (articles: 17%; CP 31%). Global level announcements and summits such as the MDGs (Millennium Development Goals) or Earth Summits (Rio+10, Rio+20) served as sustainability-related points of reference to identify trend inflections. Articles follow a consistently upward trend. The trend steadily increased with few exceptions in 2011-2012 and 2014-2015. After 2014, CP output seems to follow a U-shaped trend. Following 2015, or the SDGs announcement, the increase has no apparent trigger.

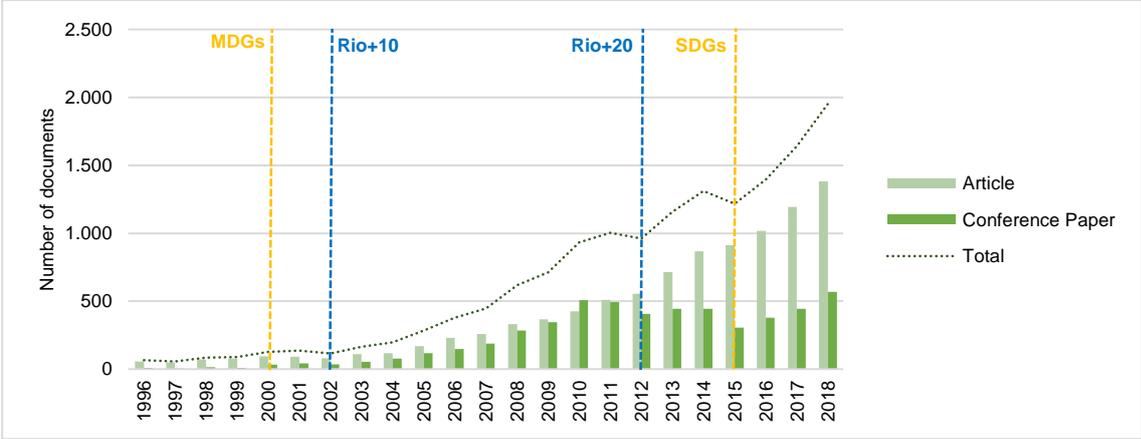

**Figure 2 Output by year and type of document for 1996-2018. Source: the authors, based on Scopus (2018) and Cortés-Sánchez et al. (2020). Note: MDGs: Millennium Development Goals; SDGs: Sustainable Development Goals.**

The fragment analysis detected subjects such as computer science/networks and communications, which are otherwise obscured when analyzing articles in isolation. Fig. 3 shows the five most productive countries by continent, normalized by the average number of researchers in R&D per million people 1996-2015. Asia is the GS leader, with China (50%), India (10%), Iran (5%), and Malaysia (3%) amassing 68% of the total output. Brazil stands for LAC with 7%. India's documents/researcher ratio was higher (10.5) than that of China (9.4). Nigeria (2.6) and Brazil (2.1) lead in their respective (sub)continent.

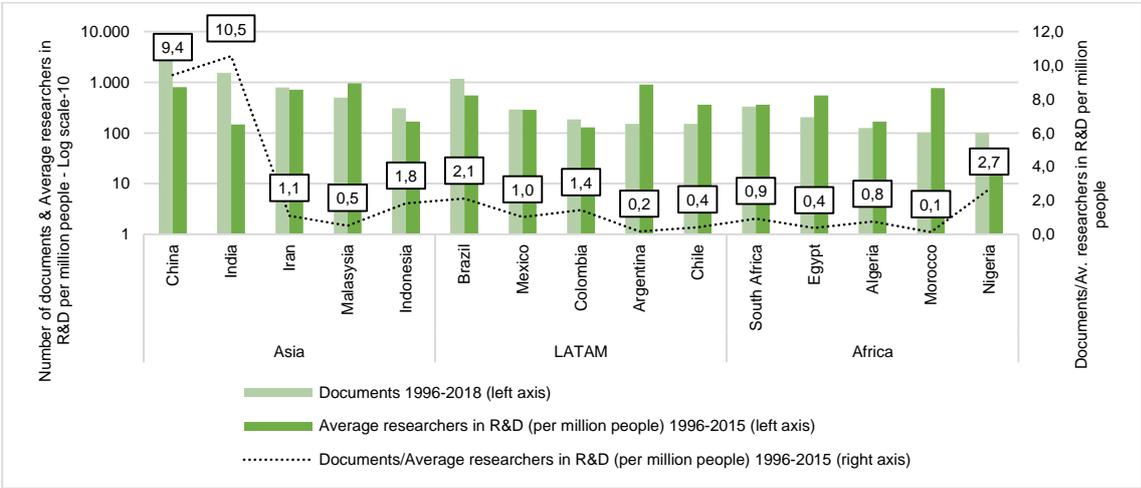

**Figure 3 Five most productive countries per continent normalized by the average number of researchers in R&D per million people 1996-2015. Source: the authors, based on Scopus (2018); The World Bank (2018); and Cortés-Sánchez et al. (2020).**

Fig. 4 shows the most productive institutions in the most productive country in each continent. All institutions are public. The Chinese universities listed are part of the C9 League, an exceptional group of elite public universities (Australian Government – Department of Education, n.d.).

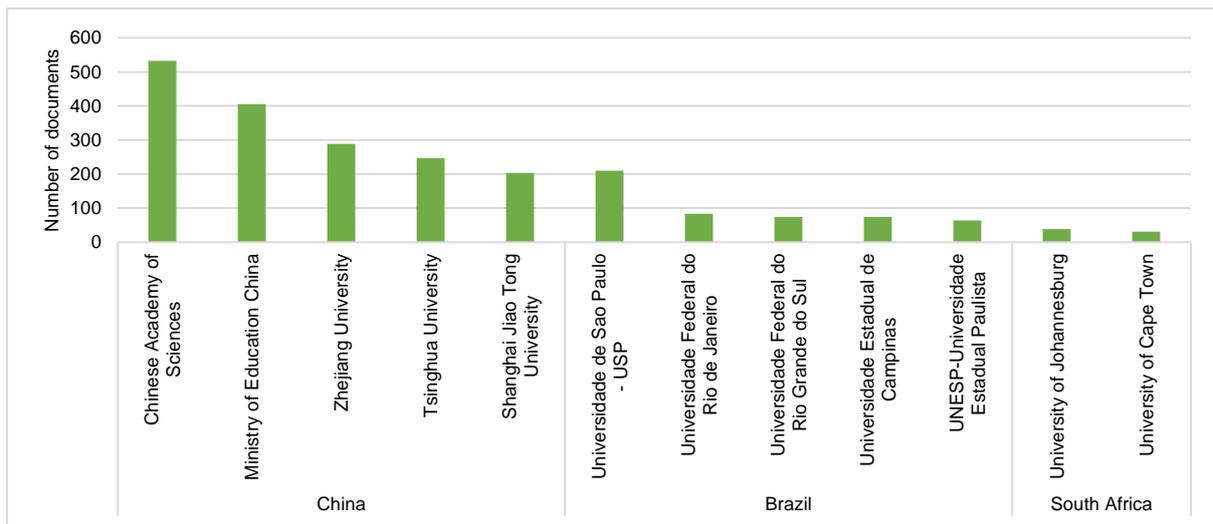

**Figure 4 Most productive institutions in the most productive country in each continent. Source: the authors, based on Scopus (2018) and Cortés-Sánchez et al. (2020).**

The country-level citation analysis used the 2,000 most cited documents. Highly cited papers provide evidence regarding transparency and cognitive and institutional differentiation and offer a threshold for domestic and international comparisons (Tijssen et al., 2002). Figs. 5 and 6 show the five most productive countries by continent for total citations and citation/document ratio for articles and CPs, respectively. For articles, China amassed 30,000+ citations and a citations/article ratio of 46.8. Nonetheless, the citations/article performance of Mexico (53.1), Chile (52.7), Nigeria (63.9), and Algeria (53.5) surpassed that of China. China also produced the highest number of total citations (3,700+). Again, however, the citation/CP performance of multiple countries surpassed that of China (4.9), such as Iran (6.2), Mexico (11.3), or South Africa (8.1).

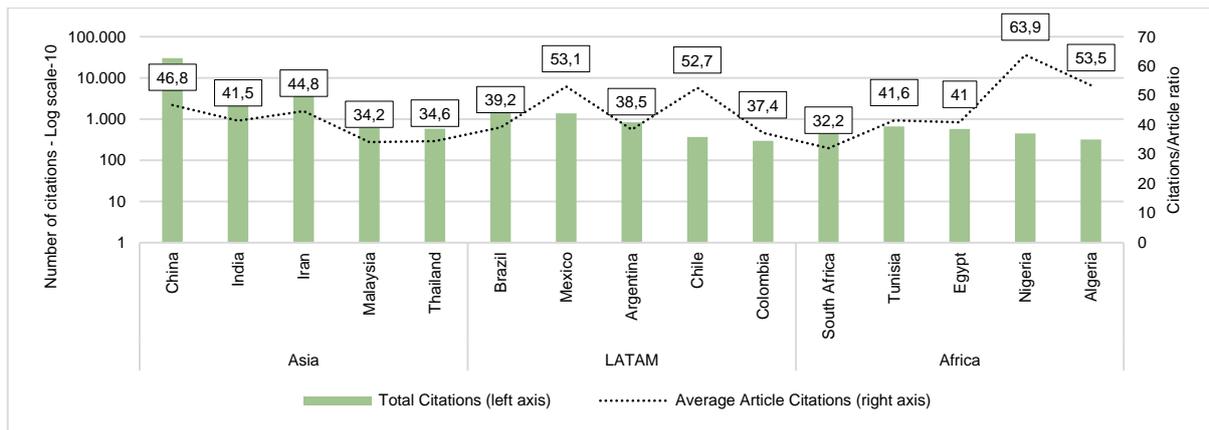

**Figure 5** Five most productive countries per continent, measured by total article citations and average article citations. Based on the 2000 most cited articles. Source: the authors, based on Scopus (2018) and Cortés-Sánchez et al. (2020).

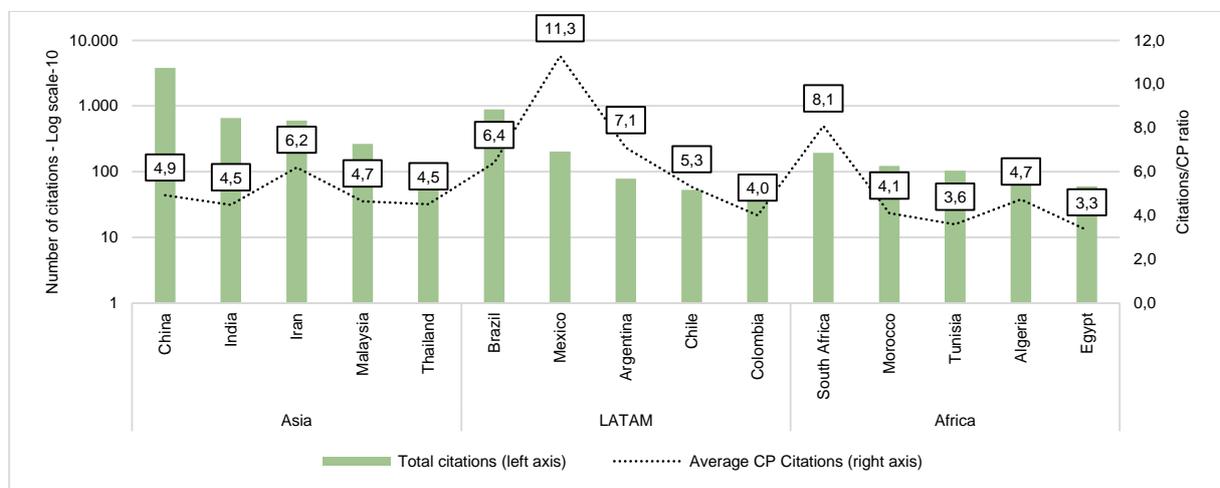

**Figure 6** Five most productive countries per continent, measured by total CP citations and average CP citations. Based on the 2000 most cited CPs. Source: the authors, based on Scopus (2018) and Cortés-Sánchez et al. (2020).

Pearson correlations between four bibliometric indices showed a positive correlation at the *p*<.01 (*n*=2,665) (i.e., SCImago journal rank (SJR); H Index; impact per publication (IPP); and source normalized impact per publication (SNIP)). Despite the correlation, bibliometric indices are not generically similar (Kosmulski, 2013). The SNIP equals the number of citations given in the present year to publications in the past three years divided by the total number of publications in the past three years, normalizing citations to correct for differences in citation practices between scientific fields (Waltman et al., 2013). That advantage is crucial given the broad sample of disciplines/subjects. The SNIP showed an r=.48 with the SJR; r=.51 with the H Index; and r=.61 with the IPP. The journal and CP with the highest SNIP were *Current Diabetes Reviews*; and *Proceedings of the Annual ACM Symposium on Theory of Computing*, respectively.

The three groups of journals and CPs used for reputational segmenting were:

- Exceptional impact-G1: top 1%, corresponds to SNIP≥2.87, *n*=50 (e.g., *Science* or *Entrepreneurship and Sustainability Issues*)
- High impact-G2: top >1% to top 10%; corresponds to 2.87>SNIP≥1.35; *n*=717 (e.g., *PLoS Biology* or *IEEE Transactions on Knowledge and Data Engineering*); and
- Middle-Low impact-G3: the remaining 89%, corresponds to SNIP<1.35; *n*=7,979 (e.g., *Freshwater Biology* or *Management Accounting Research*).

These thresholds were proposed by Tijssen et al. (2002) for highly cited papers. Figs. 7 and 8 show output by impact group and citations per document ratio. Both plots consider differences in expected life-impact: five years for articles and two for CPs (Michels and Fu, 2014). The citations per article ratio were more stable than that for CPs. The citations per article ratio fluctuated between 0.03-0.4 with a peak during 1996-2000. Citations per CP ratio fluctuated between 1.2-13.8 with a peak during the period 1998-1999. That is consistent with the statement that older documents accumulate more citations on average because they have been in the public domain for longer. Most publications of both articles and CPs are in group 3. However, since 2006 there has been an increase in article publications in group 1.

Appendix A lists the ten most-cited documents. The subjects of these articles are not restricted nor concentrated on a single topic. Topics ranged from aquatic science to geography, from agronomy and crop science to economics. All other articles belong to groups 2-3 (e.g., *PNAS, Research Policy, Macromolecules*). Authors were mostly affiliated with Chinese institutions (e.g., Ministry of Education, Zhejiang University, Shanghai Jiao Tong University). Appendix B lists the most cited articles with corporate collaboration. SciVal reported 195 (1.6%) documents with at least one corporate author. The most productive corporate-academic collaboration countries were: China (52%); India (15%); Brazil (8%); Colombia (3%); Mexico (3%); Egypt (2%); and Ghana (1%). The most active corporate organizations were public and represented three main economic sectors: energy, high-tech, and pharmaceutical sectors. They include: PetroChina (3%; Public); Guangdong Power Grid Corporation, China (3%; Public); Huawei Technologies Co., China (3%; Private); Dr. Reddy's Laboratories Ltd., India (2%; Public); Society of Petroleum Engineers

International, USA (2%; NGO); AstraZeneca, UK (2%; Private); and Petrobras, Brazil (2%; Public); and Agrosavia, Colombia (2%; Public).

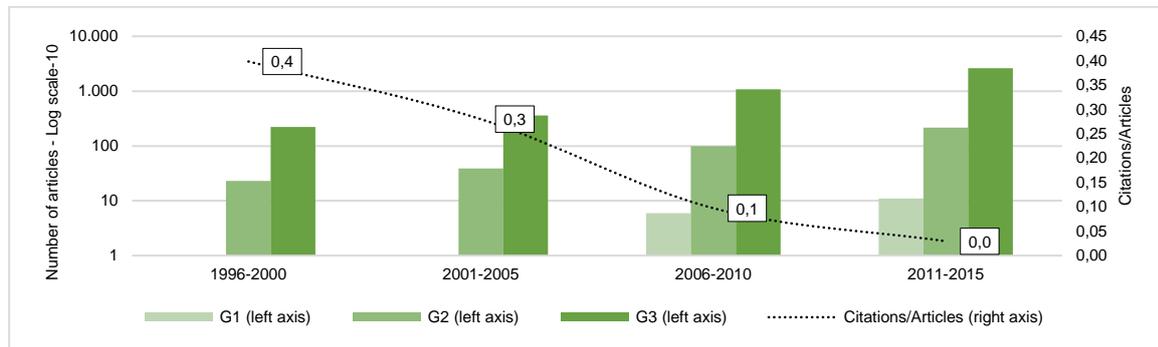

**Figure 7 Output by type of impact group and citations per article. Source: the authors, based on Scopus (2018); SCImago (2018); and the Centre for Science and Technology Studies (CWTS) (n.d.).**

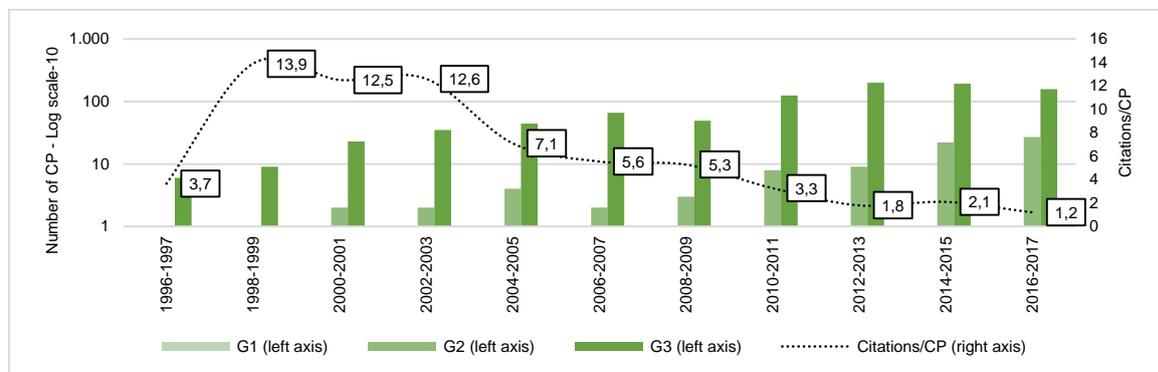

**Figure 8 Output by type of impact group and citations per CP. Source: the authors, based on Scopus (2018); SCImago (2018); and the Centre for Science and Technology Studies (CWTS) (n.d.).**

Fig. 9 shows the BCN. It is composed of 1,140 publications and 16,173 inter-publication links, with a density of 0.01. The density measures the proportion of actual links in a network divided by the potential links (Wright, 2015). The network is composed of 34 clusters. Two clusters have more than 100 publications, six clusters have between 10 and 100 publications, and 26 clusters have less than ten publications.

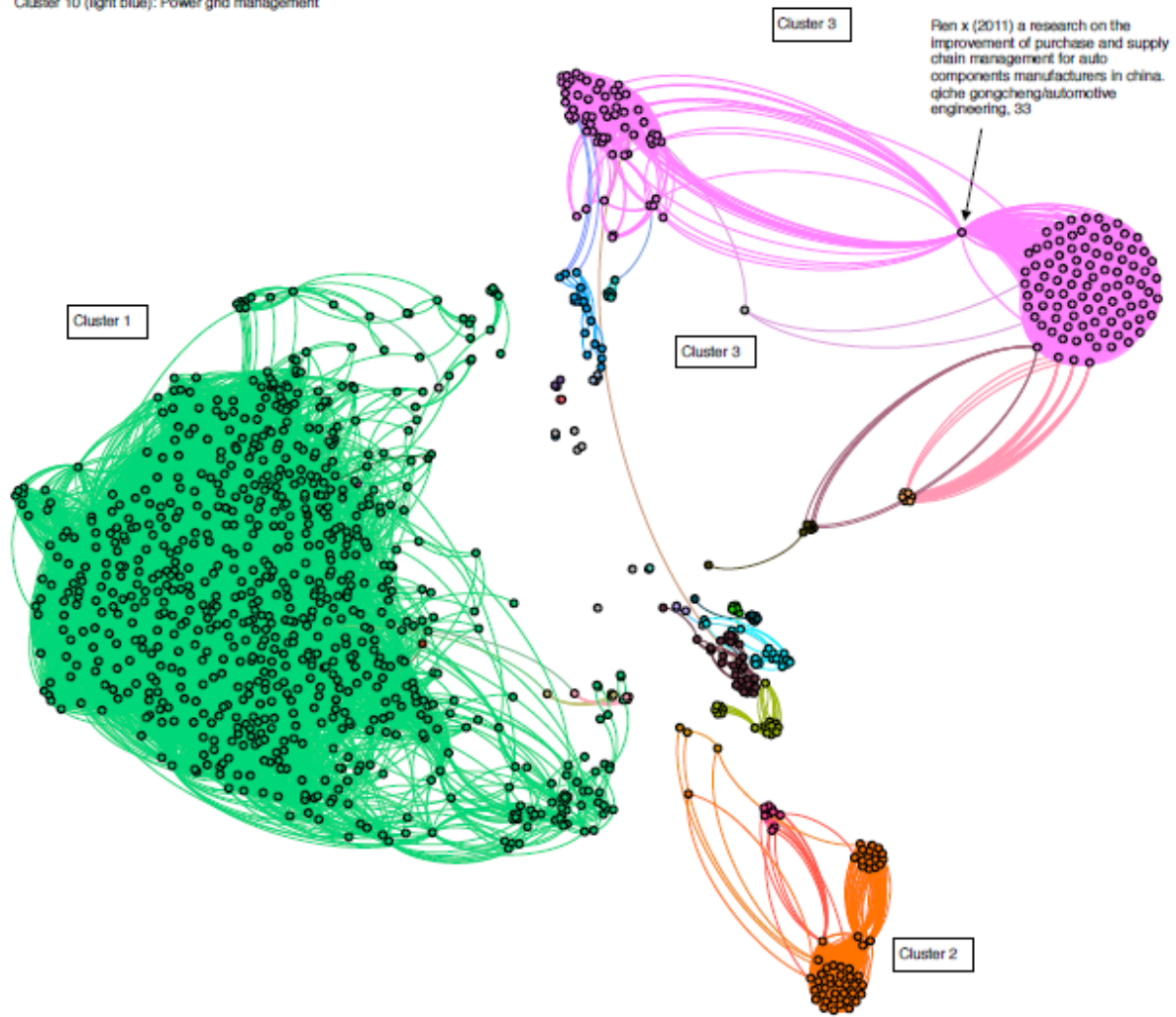

**Figure 9 Bibliographic coupling network. Source: the authors, based on Scopus (2018), processed with VOSviewer (van Eck and Waltman, 2010) and visualized with Gephi (Bastian et al., 2009).**

Network centrality measures the *importance* of an actor in terms of its relation to other actors, both in number and importance of other actors (Wright, 2015). Topics with central position (i.e., centrality) in the network are: cluster 1-*knowledge management* (density=0.02; centralization=0.09), cluster 2-*biotechnology for agriculture* (density=0.28; centralization=0.21), cluster 3-*governance of projects and power grid management in China* (density=0.21; centralization=0.18), cluster 5-*routing problems* (density=0.27; centralization=0.24), cluster 6-*sustainable farming* (density=0.11; centralization=0.09), cluster 7-*computation* (density=0.14, centralization=0.27), cluster 8-*microbiology and biotechnology* (density=0.50,;

centralization=0), and cluster 10-*power grid management* (density=0.13, centralization=0.21). The two most significant clusters are cluster 1-*knowledge management* and cluster 3-*governance of projects and power grid management in China.*

The network exhibits a high modularity structure. The paper with the most significant influence in terms of betweenness (i.e., bridging between other papers) is "*Contingencies in collaborative innovation: Matching organizational learning with strategic orientation and environmental munificence*" (Zhao et al., 2013). It examines the fit between organizational learning and contingency factors on collaborative-innovative performance in Chinese R&D firms. Its dominance in the cluster 1-*knowledge management* cluster could be related to its earlier publication year and its lack of disciplinary focus. While it analyses R&D, this is less specific and more relatable than research on supply chain management.

For this reason, knowledge management scholars in other fields can relate to its content. While there is no clear relation to sustainability, this paper is relevant for the managing of R&D staff and collaboration between R&D departments within and across companies and institutions. However, cluster 3-*governance of projects and power grid management in China* (pink) is fragmented in two (*grid management* and *projects governance*) but connected through the article by Ren (2011). Clusters 3, 5 (*routing problems*), and 10 (*power grid management*) are also connected.

The main research topics and their placement in the BCN concerning the highly cited research with corporate-academic collaboration were as follows:

- New/improved control system for voltage balancing: cluster 10-*power grid management*.
- Discovery of natural compounds and galactose utilization: clusters 2-*biotechnology for agriculture*, and 8-*microbiology and biotechnology*.
- Reuse of reverse osmosis: cluster 8-*microbiology and biotechnology*.
- Strategy for immunogenic design: cluster 8-*microbiology and biotechnology*.
- Geometric accuracy: clusters 5-*routing problems* and 7-*computation*.
- Microgrid transient dynamic: cluster 10-*power grid management*.

- Vascularization: 2-*biotechnology for agriculture*, and 8-*microbiology and biotechnology*, and
- Water purification membranes: cluster 8-*microbiology and biotechnology*.

Clusters 8-*microbiology, biochemistry, and biotechnology*; and 2-*biotechnology for agriculture* were the focus of corporate-academic collaborations. *Water purification membranes* was the only topic of the academic-corporate collaboration also spotted in the trending key-phrase analyses.

Collaboration, measured by the number of authors and extent of national/international collaboration, varied substantially for articles and CPs. Single-authored documents are the exception. There is a higher international collaboration for articles than for CPs and exclusively national collaboration is apparent in the case of articles. SciVal's sample reported 6.7% single-authored articles, 31.7% of articles involving national collaboration only, and 27.1% involving international collaboration. Among CPs, 12.3% were single-authored, 22.8% were national collaboration only, and 14.3% involved international collaboration.

Individual-level co-authorship insights were unsupportive due to the high number of researchers affiliated with Chinese institutions. Fig. 10 shows the CAN of countries with at least five co-authored articles or CPs among the 2,000 most cited documents. The frequency of co-authored documents is higher for North-South collaboration than for South-South collaboration. LAC is the GS region with the highest collaboration within its countries and between other GS regions. China has the highest number of documents co-authored with developed countries (articles with the US: 114 and the UK: 33; CP with the US: 44 and the UK: 14) followed by Brazil (articles with the US: 20; CP with the US: 11 and Germany: 9) and South Africa (articles with the US: 13). The South-South collaboration distribution was as follows:

- LAC: articles: Brazil-Argentina=3; CP: Brazil-Chile=2.
- Africa: articles: South Africa-Malawi=2.
- LAC-Asia: articles: Brazil-India=2; CP: Brazil-India=2; China-Argentina=2; Brazil-China=1.
- LAC-Africa: articles: Brazil-Kenya=1.
- Africa-Asia: articles: South Africa-Japan=1; CP: South Africa-China=1.

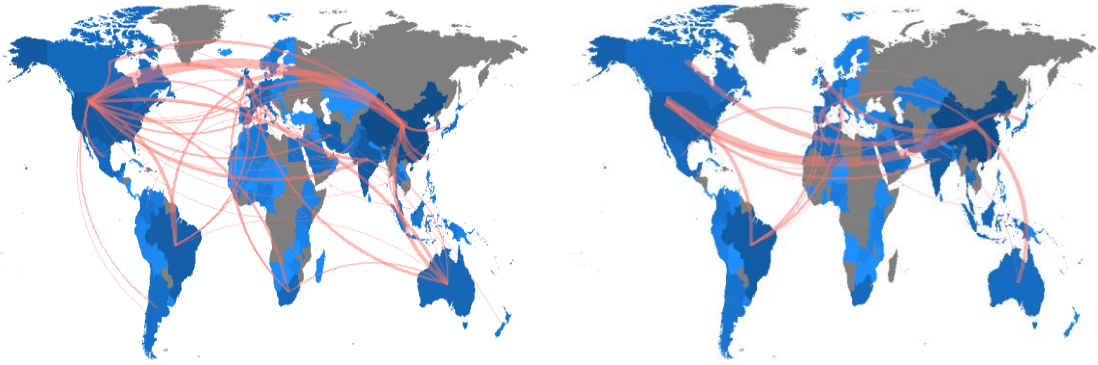

**Figure 10** International collaboration (left: articles; right: CPs). Source: the authors, based on Scopus (2018); Cortés-Sánchez et al. (2020).

## 5    Discussion

Global research and iS-GS are jointly growing, the former as part of the latter. There is a similar average percent growth between iS-GS (18%), and the overall percent growth of Chinese research (15%) 1996-2018 as the overall output leader (SCImago, 2018). iS-GS output after global level announcements and summits such as the SDGs or the Earth Summits increased steadily with few exceptions. In this line, evidence shows that an improvement in environmental governance positively affects research production in the environmental sciences and ecology (Dragos and Dragos, 2013). Before 2010, global research related to the SDGs amounted to fewer than 100 published papers per year but had grown to more than 500 papers per year by 2016 (Nakamura et al., 2019).

Significant regional publishing gaps persist (Cortés-Sánchez et al., 2020). In Asia, China published 25 times more documents than Indonesia; in LAC, Brazil published eight times more than Chile; in Africa, South Africa published three times more than Nigeria. Explanatory factors for author productivity, institutional productivity, and national productivity include author gender – but not exclusively (van Arensbergen et al., 2012), incentives and collaboration dynamics (Defazio et al., 2009), grant funding (Jacob and Lefgren, 2011), culture (Inönü, 2003); and population and national public/private investment in R&D (King, 2004). Regarding the last, this study concurs with the findings of Prathap (2017) on the relationship between R&D and scientific wealth (i.e., production and impact) and the concentration of such wealth in a few premier public institutions. The 12 most productive institutions' total publications accounted for 15% of the iS-GS output for the whole GS.

As an exceptional case, the scientific wealth (i.e., production and impact) of China could be significantly driven by the population and national public/private investment in R&D (King, 2004). China's R&D investment followed that of the USA with USD 400 billion. Furthermore, in 2019, China's spending on science and research funding reached 2.5% of its total GDP, as it seeks to catch up on the technology front (Ng and Cai, 2019). The five most productive institutions in terms of net-output were public, typically universities.

China leads both output and citations, but several countries such as Mexico, Nigeria, Argentina, or Algeria, surpassed its citations/document ratio. This pattern highlights the geographically diverse roots of sustainability research (Bettencourt and Kaur (2011). China's average citations per document (9.4) is below the global average (11.8) (Huang, 2018), yet higher than the LAC average (5.2) on innovation related-research (Cortés-Sánchez, 2019). Perverse incentives may partly explain China's inferior citation performance: researchers in China do not secure research funding as principal investigators unless they publish at least half a dozen articles indexed in the Science Citation Index. Journal reputations aside, a researcher could receive between US$900 and US$10,000 for each article published (Huang, 2018). It is essential to clarify that this is a failure of the research system rather than a cultural or national problem (Edwards and Roy, 2016). Both social and individual costs must be considered in designing and implementing improved incentive schemes (Stephan, 2012).

Countries like Mexico, Nigeria, and South Africa are increasing their scientific wealth, mainly STEM. Research in Computer Science, led by the largest university in Mexico and LAC (*Universidad Nacional Autonoma de Mexico*) and their National Council of Science and Technology (CONACYT), is generating higher citations per paper average than the world's 100 most productive institutions (Uddin et al., 2015). In Africa, South Africa and Egypt produce most of the output (Pouris and Pouris, 2009). STEM research in Nigeria is flourishing but lacks international visibility (Usman and Ewulum, 2019). Both Nigeria and South Africa dominate research output and use of academic-corporate collaboration research in the Sub-Saharan region (Zavale and Langa, 2018). That aligns with growth expectations for the GS in research production (Finardi, 2015).

Previous research found that iS-GS publications in highly reputable outlets were inclined towards STEM rather than business and management (e.g., electrical and electronic engineering; agronomy; chemical engineering; and computer science) (Cortés-Sánchez et al., 2020). Compared with lists of reputable journals on sustainability (Zhu and Hua, 2017) and innovation (Merigó et al., 2016) such as *Science* or *Energy Policy,* and *Research Policy* or *Strategic Management Journal*, only the *Journal of Cleaner Production* and *Energy* were among the top-ten journals identified with cumulative iS-GS growth. Two factors partly explain variations in citation dynamics and a journal's reputation. Social sciences have tended to be more locally oriented and linguistically fragmented than natural sciences (Dyachenko, 2014). Also, at least in computer science reviewing, evidence suggests a bias in favor of authors from English-speaking countries and affiliated with prestigious institutions (Walker et al., 2015).

There are few resemblances between the current BCN and those previously documented. Cognitive similarities and differences between fields involved in iS-GS are diametrically distinct from those related to the SDGs and open innovation. Clusters 2- *biotechnology for agriculture* and 6-*sustainable farming* are the only ones related to the *sustainable* cluster found in Nakamura et al's (2019) study. Conversely, the clusters most related to innovation in management for sustainability according to Nakamura et al. (2019) (i.e., *green supply chains and management; manufacturing/remanufacturing systems; cost analysis and optimization models for waste management and recycling*) were not evident in our exercise.

The central clusters identified here diverge from those identified as central by Nakamura et al. (2019) (i.e., *maternal, newborn, and child morbidity and mortality; ecosystems services and adaptations for sustainability; global, regional, and national health surveys; diagnosis and management of tuberculosis; substance abuse and longevity*). Further, clusters 3-*governance of projects and power grid management in China* and 6-*sustainable farming* are related to the cluster of *local development change in the relations between local civil communities and their governing bodies* identified by van der Have and Rubalcaba (2016) in studying open innovation. Since there is a systematic relation between SDGs, it is improper to state that iS-GS contributes only to a particular SDG. However, central clusters were closely related to four SDGs: industry, innovation, and infrastructure; decent work and economic growth; affordable and clean energy; and zero hunger.

The bibliographic database may be a factor in the differences found between these BCN results. The WoS bibliographic database was used for both studies on SDGs and open innovation. Compared with WoS, Scopus has broader journal coverage and greater social sciences coverage (Gavel & Iselid, 2008; Mongeon & Paul-Hus, 2016). In sum, mapping results offer a deeper understanding and categorization of the relatedness and knowledge flows between fields and show researchers, universities, and governments how their current research relates to other research activity (Guevara et al., 2016).

Nevertheless, in a broader aspect, the Hierarchy of Sciences hypothesis could shed light from another angle. The hypothesis states that while some sciences/disciplines studying simple phenomena (i.e., cells' functioning) will tend toward consensus, others studying complex phenomena (i.e., human collective behavior) will tend toward dissension (Fanelli and Glänzel, 2013). Thus, business and management will tend towards dissension and STEM consensus. That could help understand the highly BCN modularity (i.e., the largest cluster related to - *knowledge management*, followed by *biotechnology for agriculture,* both in different and separate clusters, the former towards dissension and the latter toward consensus*).*

In the GS, the corporate-academic collaboration is still in its infancy, and its diversity is still low (i.e., focused on three economic sectors: energy, high-tech, and pharmaceutics). In China, public firms will continue to influence the corporate-academic research agenda because they participate in over half of China's research output. In Brazil, Petrobras has a significant presence in energy-related research (Cortés-Sánchez, 2019). The evidence suggests several strategies and determinants for consolidating and increasing university-corporate collaboration (Bodas Freitas et al., 2013; Maietta, 2015; Soh and Subramanian, 2014; Sutz, 2000). First, it is vital to promote information exchange between universities and industries to build capacities for problem-solving and knowledge on both sides. A focus on a particular branch of enterprise could strengthen the pool of competence between university faculty and firms. Moreover, universities could adopt a differentiated governance approach for different sized firms and offer degrees valuable to neighboring firms.

Teams domain the global production of knowledge (Wuchty et al., 2007). Hence, single-authored documents are uncommon. GS collaboration is more frequent with the North than within the South. Research on the SDGs in Africa, the Arab States, and LAC, is more frequently linked to authors from Europe or North America than from the South (Nakamura et al., 2019). Half of the publications of the 500 most-cited African researchers include authors from non-African institutions. The US is the preferred collaboration partner for Asian countries in physics and life sciences (Arunachalam and Doss, 2000). Adams (2012) also noted that LAC is an emerging network orbiting Brazil, while Africa is developing three different language-similar networks in southern Africa, West Africa (French-speaking countries), and East Africa (English-speaking countries). Further, the most influential innovation scholars are (were) affiliated with leading US or European universities, which also generate 86% of the knowledge in the field (Fagerberg et al., 2012).

As a final point, tensions between North-South are emerging due to differences in research priorities and funding. For instance, NASA –a western funding agency– funds astronomy or astrophysics research in Africa, while local researchers and institutions cast doubt on the local importance/priority of that research (Kozma and Calero-Medina, 2019). Both historically rooted and current inequalities may explain the North-South scientific wealth gap (Jentsch and Pilley, 2003). Differences in research roles persist, where the North remains a provider of ideas and funding, while the South is often relegated to the receiver of funds and data quarry (Jentsch and Pilley, 2003).

# 6   Conclusion

This study's aimed to conduct a comprehensive bibliometric outlook on the output, impact, and structure of the research on innovation for sustainability in the iS-GS. In terms of output and impact, iS-GS is growing at an 18% rate and has been driven by global announcements and summits since environmental governance improves research production. National level institutions should examine inter/intra countries' research output gaps now that leading output countries can produce up to 25 times that of the 2$^{nd}$ most productive country. Explanatory factors for that range from gender to population and national public/private investment in R&D. Geographical diverse roots and decisive support of public institutions are driving forces for other countries than China to increase scientific wealth in the GS (e.g., México, Nigeria, South Africa). GS researchers are publishing in very few journals at a cumulative growth pace. That could be partly explained by the bias in favor of authors from English-speaking countries and affiliated with prestigious institutions, few of them in the GS.

In terms of structure, iS-GS clusters are peripherical compared to those vertebral for SDGs research (e.g., maternal, newborn, and child morbidity and mortality). Nevertheless, iS-GS clusters were related to four SDGs: industry, innovation, and infrastructure; decent work and economic growth; affordable and clean energy; and zero hunger. The high BCN modularity could be explained by the dissension vs. consensus of the business and management and STEM research, supported by the Hierarchy of Sciences hypothesis. The academic-corporate collaboration has been focused on just three sectors, energy, high-tech, and pharmaceutics. There is a lack of South-South collaboration since North-Sout collaboration is more robust and, as a collateral effect, is causing tensions between local and international funders.

A statement that needs further support is that institutional governance's environmental or innovation-related improvement is positively related to scientific wealth, not only output. Accordingly, The reasons for the limited amount of South-South collaboration is worth exploring further. Improved incentives, both material and intangible, could also help to expand former and further academic-academic and corporate-academic collaboration opportunities.

Further research could compare bibliographic data from other databases (i.e., WoS; Dimensions; Google Scholar; Microsoft Academic), and/or include another type of literature (i.e., patents). Alternative metrics (e.g., altmetrics) might highlight the incidence of iS-GS for a broader audience beyond the academic niche. In-depth comparative perspectives between the GS and the Global North could outline shared or differentiated research resources, capabilities, and priorities. Finally, an analysis of national or organizational research policies and incentives could help explain historical and local particularities and help design corporate-academic collaborations oriented towards increasing research output and impact.

**Acknowledgments**

[Pending]

# Appendices

**Appendix A Top-10 most cited documents**

| # | Citations | 1st author last name | 1st author affiliation | Year | Title | Journal | SJR | H index | IPP | SNIP | SNIP Group |
|---|---|---|---|---|---|---|---|---|---|---|---|
| 1 | 1136 | Ju | Key Laboratory of Plant and Soil Interactions, Ministry of Education, China | 2009 | Reducing environmental risk by improving N management in intensive Chinese agricultural systems | PNAS | 5,6 | 699,0 | 9,6 | 2,4 | 2 |
| 2 | 1059 | Hobbs | School of Environmental Science, Murdoch University, Australia | 2006 | Novel ecosystems: Theoretical and management aspects of the new ecological world order | Global Ecology and Biogeography | 3,5 | 127,0 | 0,9 | 0,6 | 3 |
| 3 | 816 | Poff | Environmental Flow Specialists, Inc., United States | 2010 | The ecological limits of hydrologic alteration (ELOHA): A new framework for developing regional environmental flow standards | Freshwater Biology | 1,7 | 139,0 | 2,0 | 1,3 | 3 |
| 4 | 540 | Stilgoe | University of Exeter Business School, Department of Science and Technology Studies, United Kingdom | 2013 | Developing a framework for responsible innovation | Research Policy | 3,4 | 206,0 | 1,1 | 1,5 | 2 |
| 5 | 423 | Hilborn | Sch. of Aquatic and Fishery Sciences, University of Washington, United States | 2004 | When can marine reserves improve fisheries management? | Ocean and Coastal Management | 1,0 | 70,0 | 0,4 | 0,5 | 3 |
| 6 | 400 | Owen | University of Exeter Business School, United Kingdom | 2012 | Responsible research and innovation: From science in society to science for society, with society | Science and Public Policy | 0,7 | 55,0 | 0,4 | 1,0 | 3 |
| 7 | 348 | Cassman | Agronomy Department, United States | 1998 | Opportunities for increased nitrogen-use efficiency from improved resource management in irrigated rice systems | Field Crops Research | 1,7 | 127,0 | 1,1 | 1,2 | 3 |

| 8 | 288 | Shen | Institute of Power Electronics, Zhejiang University, China | 2008 | An improved control strategy for grid-connected voltage source inverters with an LCL filter | IEEE Transactions on Power Electronics | 2,5 | 222,0 | 0,5 | 2,0 | 2 |
| 9 | 287 | Li | School of Chemistry and Chemical Engineering, Huazhong University of Science and Technology, China | 2011 | A new strategy to microporous polymers: Knitting rigid aromatic building blocks by external cross-linker | Macromolecules | | 2,2 | 288,0 | 3,1 | 1,9 | 2 |
| 10 | 271 | Zhou | National Key Laboratory of Micro/Nano Fabrication Technology, Shanghai Jiao Tong University, China | 2012 | Photo-Fenton reaction of graphene oxide: A new strategy to prepare graphene quantum dots for DNA cleavage | ACS Nano | 6,2 | 310,0 | 2,9 | 1,0 | 3 |

Source: the authors' based on Scopus (2018).

**Appendix B Top-ten most cited articles in academic-corporate collaboration**

| Citations | Leading author | Year | Title | Source | Corporation | Academic Ins. |
|---|---|---|---|---|---|---|
| 123 | Fan, S. | 2015 | An improved control system for modular multilevel converters with new modulation strategy and voltage balancing control | IEEE Transactions on Power Electronics | Siemens | Huazhong University of Science and Technology |
| 112 | Zhou, J. | 2011 | Direct synthetic strategy of mesoporous ZSM-5 zeolites by using conventional block copolymer templates and the improved catalytic properties | ACS Catalysis | SINOPEC, CAS - Shanghai Institute of Ceramics | Chinese Academy of Sciences |
| 102 | Yang, W.-Z. | 2012 | Panax quinquefolium and Panax notoginseng to characterize 437 potential new ginsenosides: A strategy for efficient discovery of new natural compounds by integrating orthogonal column chromatography and liquid chromatography/mass spectrometry analysis: Its application in Panax ginseng | Analytica Chimica Acta | Agilent Technologies | Peking University |
| 97 | Badruzzaman, M. | 2009 | Innovative beneficial reuse of reverse osmosis concentrate using bipolar membrane electrodialysis and electrochlorination processes | Journal of Membrane Science | MWH Global, Inc. | University of Illinois at Urbana-Champaign |
| 95 | Hong, K.-K. | 2011 | Unravelling evolutionary strategies of yeast for improving galactose utilization through integrated systems level analysis | Proceedings of the National Academy of Sciences of the United States of America | Novo Nordisk A/S, CJ CheilJedang Corporation | Soochow University, Technical University of Denmark, Chalmers University of Technology |
| 72 | Ping, L.-H. | 2013 | Comparison of viral env proteins from acute and chronic infections with subtype C human immunodeficiency virus type 1 identifies differences in glycosylation and CCR5 utilization and suggests a new strategy for immunogen design | Journal of Virology | Leading: Bristol-Myers Squibb | University of Georgia |
| 65 | Malhotra, R. | 2011 | Improvement of geometric accuracy in incremental forming by using a squeezing toolpath strategy with two forming tools | Journal of Manufacturing Science and Engineering, Transactions of the ASME | Ford Motor | Northwestern University - Indian Institute of Technology, Kanpur |
| 58 | Kamel, R.M. | 2013 | Three control strategies to improve the microgrid transient dynamic response during isolated mode: A comparative study | IEEE Transactions on Industrial Electronics | Hitachi | Assiut University - Tokyo University of Agriculture and Technology |
| 57 | Egaña, J.T. | 2009 | Use of human mesenchymal cells to improve vascularization in a mouse model for scaffold-based dermal regeneration | Tissue Engineering - Part A | Roche Diagnostics GmbH Technical | University of Munich - University of Lübeck - Universidad de Chile - Technische Universität Dresden |
| 55 | La, Y.-H. | 2011 | Bifunctional hydrogel coatings for water purification membranes: Improved fouling resistance and antimicrobial activity | Journal of Membrane Science | IBM - King Abdulaziz City for Science and Technology | University of Texas at Austin |

Source: the authors' based on Scopus (2018)